# A Study of Time Variations of UV Continuum and Emission Lines in 3C 390.3

Wei Zheng

Center for Astrophysical Sciences, The Johns Hopkins University

Baltimore, MD 21218–2695

Electronic mail: zheng@pha.jhu.edu

## ABSTRACT

Based on the *IUE* SWP spectra obtained between 1978 and 1994, the variations of the UV continuum and the Ly$\alpha$ and C IV emission lines of the broad–line radio galaxy 3C 390.3 are studied. The UV continuum between 1220 and 1990 Å has varied considerably. In 1991 August, the flux at 1360 Å reached a maximum level of $1.3 \times 10^{-14}$ erg cm$^{-2}$ s$^{-1}$ Å$^{-1}$, $\sim 27$ times as high as its minimum level. At 1850 Å, the continuum varies by a factor of $\sim 12$. Significant variations are frequently observed in periods as short as one month.

The total flux of Ly$\alpha$ and C IV emission varies by a factor of $\sim 5$. Their profiles are fitted with one narrow component, unresolved at the instrumental resolution of FWHM $\approx 2000$ km s$^{-1}$, and three broad components (FWHM $\approx 6000$ km s$^{-1}$). The narrow component varies by a factor of $\sim 2.5$. The broad components on the blue and red wings of the Ly$\alpha$ and C IV emission profile, which correspond to the double humps in the H$\alpha$ and H$\beta$ emission feature, display a close correlation to the varying continuum, with a time lag of $\sim 45$ days.

The blue Ly$\alpha$ component has been stronger than the red component except for several epochs in 1980 and 1991. The C IV/Ly$\alpha$ flux ratio is $\sim 0.4$ for the narrow component. For the broad components, this ratio varies considerably around an average value of unity. In 1991 March, the blue Ly$\alpha$ component was stronger than the red, while for C IV the red component was stronger. Such differential variations do not match the accretion disk characteristics. The findings suggest that the narrow– and broad–line region in 3C 390.3 may be associated with a radio jet.

*Subject headings:* galaxies: active — galaxies: individual (3C 390.3) — ultraviolet: galaxies

## 1. INTRODUCTION

It is now well known that most active galaxies are variable, and emission lines in their spectra vary in step with the underlying continuum. As the line–emitting region extends a range of distance from the central power source, the correlation between emission lines and the continuum serves as a tool in mapping the geometrical and dynamical structure of the broad–line region (BLR, Peterson 1988, 1993 and references therein). In recent years, monitoring campaigns of high temporal frequency have considerably enhanced our knowledge of the BLR. The extensive monitoring studies of Seyfert galaxy NGC 5548 (Clavel et al. 1991; Peterson et al. 1991, 1992, 1994; Korista et al. 1995) with the International Ultraviolet Explorer (*IUE*), the Hubble Space Telescope, and ground–based telescopes find that the scale of continuum variations decreases toward longer wavelengths and that the variations of the optical continuum closely follow those in the UV band without a detectable delay ($\Delta t \sim 1$ day). It has also been found that the bulk of the variable BLR lies within two lt–weeks of the central power source, and that high–ionization lines respond faster than low–ionization lines to changes in the underlying continuum. Other objects that have also been studied extensively during multi–band monitoring campaigns include: NGC 4151 (Clavel et al. 1990; Edelson et al. 1994), NGC 3783 (Reichert et al. 1994, Stirpe et al. 1994), and Fairall 9 (Rodríguez–Pascual et al. 1996). The results of these studies confirm that the pattern of variations found in NGC 5548 is representative of Seyfert galaxies in general.

As the observed emission–line profiles are the result of a bulk velocity distribution around the central power source, the blue and red wings and the core of a profile should exhibit different time lags with respect to the varying continuum. At the current temporal sampling rates and data signal–to–noise levels, such a difference has not been clearly established. The recent study of the velocity–dependent response of the C IV emission line in NGC 5548 finds no evidence that the BLR kinematics involve predominantly radial motions (Korista et al. 1995). Neither infall or outflow is indicated. This result agrees with an early claim (Clavel et al. 1991) of random cloud motions in NGC 4151, with higher velocities close to the central source, and conflicts with the finding of Crenshaw & Blackwell (1990) that the velocity field in NGC 5548 is gravitational infall. In general, the results regarding the velocity field around the central power source are still ambiguous.

While most of the objects that have been frequently monitored are radio–quiet Seyfert galaxies, the emission lines in the following radio galaxies and quasars are known to vary: 3C 382 (Yee & Oke 1981; Tadhunter et al. 1986), 3C 273 (O'Brien, Zheng & Wilson 1989; Edelson, Krolik, & Pike 1990) and 3C 120 (Gondhalekar 1992). Of special interest is the broad–line radio galaxy 3C 390.3 (z=0.0561; Osterbrock, Koski, & Phillips 1975), a prototype for very broad, double–peaked emission–line profiles. Such profiles draw particular attention as they may be the signature of a relativistically rotating disk (Pérez et al. 1988). The object's high variability in the continuum as well as in the emission lines has been well–known (Barr et al. 1980; Yee & Oke 1981; Netzer 1982; Barr, Willis, & Wilson 1983; Penston & Pérez 1984). The narrow emission lines in 3C 390.3 are formed in a compact region of moderate density ($\sim 10^6$ cm$^{-3}$, Osterbrock et al. 1975) and exhibit considerable variations within several years (Clavel & Wamsteker 1987; Zheng et al. 1995). A size of several lt–yr for narrow–line regions (NLR) is smaller than the conventional estimates by more than two orders of magnitude. The complex emission–line profiles in 3C 390.3 suggest a particularly rich structure of the line–emitting region. The radio structure in 3C 390.3 displays a superluminal velocity of $\sim 4c$ for the Hubble constant H = 50 km s$^{-1}$Mpc$^{-1}$ (Alef et al. 1988). Because of the radio jet and the superluminal velocity, radial motion may play an important role in the line–emitting region in this object.

The very broad line profiles (FWHM $\approx 10^4$ km s$^{-1}$) in 3C 390.3 suggest a highly stratified velocity field in the BLR, making it an ideal candidate to study the gas kinematics. In this paper, the UV continuum, emission–line profiles and, in particular, the pattern of velocity–dependent line variations are studied. The advantage of such a study is the long time span that enables detection of dramatic variations. However, studies of archival data often suffer from poor temporal sampling and miss the details of the variations. 3C 390.3 is currently a target of extensive *IUE* monitoring, and the results of this archival study may provide clues to the pattern of variations in this object.

## 2. DATA ANALYSES

43 archival *IUE* SWP spectra were retrieved from the NASA National Space Science Data Center in the *MELO* form. The spectra form a complete set of the SWP data for 3C 390.3 between 1978 November 21 and 1994 June 17. Several spectra contain spike pixels that apparently arise from cosmic ray events. These spectra have been re–extracted from the line–by–line images using the standard *IUE* data reduction procedures available in the *IUE* Regional Data Analysis Facility. The spectral analyses have been carried out using the task `specfit` (Kriss 1994) that minimizes $\chi^2$ in fitting the continuum and line features. As *IUE* spectra do not provide errors, we assign an arbitrary error of $10^{-8}\sqrt{f}$ for each pixel, where $f$ is the flux in units of erg cm$^{-2}$ s$^{-1}$ Å$^{-1}$. The $\chi^2$ values thus reflect the progress of fitting attempts but do not indicate goodness of the fit. All the errors presented in this paper are related to the fitting process only. Systematic errors associated with the instrument and observations are not taken into account.

The fitting process consists of two steps. First, the underlying continuum is fitted with a power law in the observed wavelength windows of 1335–1380, 1407–1447, 1500–1570, and 1820–1890 Å. In the spectrum SWP3478, the wavelength region between 1820 and 1890 Å is not used because of an apparent abnormality in this window. The interpolated fluxes of the fitted power law at 1360 and 1850 Å and 1$\sigma$ errors are presented in Table 1. After fitting the continuum with a power law from the selected wavelength windows, the underlying continuum is fixed. The second step is to fit the Ly$\alpha$ and C IV lines in the wavelength windows of 1225–1338 and 1560–1680 Å. The continuum and line fittings are separated on account of the very large equivalent widths of the emission lines. In trial fittings that include the continuum and lines, the prominent emission lines often cause an apparent deviation of the fit from the underlying continuum. The two–step fitting process produces more satisfactory fits as judged by eye.

The profiles of UV emission lines in 3C 390.3, unlike their optical counterparts, are dominated by a prominent central component. A critical step in the fitting process is to estimate the narrow–component width in the *IUE* spectra. Under the best conditions, the width of the point spread function of the SWP camera is $\sim 5$ Å at the wavelength of Ly$\alpha$ (Cassatella, Barbero, & Benvenuti 1985). The real resolution is likely to be lower because of the large aperture used. In the low states in which the broad–line components virtually disappear, the narrow–line component is measured to be FWHM $\sim 9$ Å for Ly$\alpha$ and $\sim 8$ Å for C IV. To check this approximation, the *IUE* spectra of I ZW 1, a narrow–line Seyfert galaxy, have been studied, and the measurements of the Ly$\alpha$ and C IV line widths of I ZW 1 agree with those of 3C 390.3 to within 10%. Therefore a narrow–line width of 9 Å ($\sim 2100$ km s$^{-1}$) should be a reasonable approximation for the Ly$\alpha$ emission feature.

In their analyses of the *IUE* spectra of 3C 390.3 between 1978 and 1986, Clavel & Wamsteker (1987) adopt a narrow Gaussian component with a fixed FWHM of 8 Å and a broad Gaussian component for both the Ly$\alpha$ and C IV profiles. However, each *IUE* spectrum has a slightly different resolution, and a narrow component with a fixed width sometimes results in apparent residuals. The new fitting process allows the widths of narrow components to vary slightly ($9 \pm 1$ Å for Ly$\alpha$). Due to larger line widths, the fluxes of the fitted narrow components may be slightly higher than those in their work.

Because of the complex profiles of emission lines in 3C 390.3, three broad Gaussian components are assumed for the Ly$\alpha$ emission feature: central, blue and red. They are set to the same line width. The peak wavelength of the central broad component is allowed to vary between $-500$ and $+500$ km s$^{-1}$ with respect to the narrow peak. The peak wavelength of the blue and red component are allow to vary between $+4500 \pm 1000$ km s$^{-1}$ and $-3500 \pm 1000$ km s$^{-1}$, respectively. Such an assumption is made to match the double humps in the H$\alpha$ and H$\beta$ emission profile (Veilleux & Zheng 1991; Zheng, Veilleux, & Grandi 1991) as well as the humps that can be seen in some *IUE* spectra (Figure 4). A typical line width of the fitted broad components is $\sim 5000$ km s$^{-1}$. The lower flux limit for each component is set to zero. In Table 2 and 3 there are several terms with zero flux, they represent an upper limit to the flux of $\approx 3 \times 10^{-14}$ erg cm$^{-2}$ s$^{-1}$.

The widths of the C IV components are set to be the same as the respective Ly$\alpha$ components (in units of km s$^{-1}$), and the centroid wavelength of each component is fixed to that of its Ly$\alpha$ counterpart, according to the ratio of their rest wavelengths. The fitted components of the Ly$\alpha$ and C IV emission features are presented in Table 2 and 3, respectively.

## 3. RESULTS

### 3.1. Continuum

In terms of year–to–year variations, the continuum flux level has declined between 1978 and 1984. As measured at 1360 Å, it reached a minimum of $5 \times 10^{-16}$ erg cm$^{-2}$ s$^{-1}$ Å$^{-1}$ in mid–1984. There was then a considerable increase between 1984 and 1985. Between 1985 and 1991 the continuum flux gradually increased, reaching a maximum value of $1.35 \times 10^{-14}$ erg cm$^{-2}$ s$^{-1}$ Å$^{-1}$ in 1991 November. Since 1992 it has been at a moderate level. The variation scale of the flux at 1360 Å is a factor of $\sim 27$, one of the highest among active galaxies. In addition to this general trend, short–term flux variations are frequent, as shown in Table 1 and Figure 1. A dramatic variation took place between 1992 January (SWP43643) and February (SWP43951) when the continuum level varies nearly twofold within 27 days.

The variation scale of flux at 1850 Å of $\sim 12$ suggests a decreasing variability at longer wavelengths. The averaged power–law index $\beta$ ($f_\lambda \propto \lambda^\beta$) is $-0.8 \pm 0.6$ at the minimum level, and $-1.5 \pm 0.3$ at the maximum. These averaged values are calculated based on the 3 lowest and highest flux levels, respectively. Such a pattern of wavelength–dependent variations is common in active galaxies (Edelson et al. 1990). If the UV continuum of AGN arises mainly in thermal emission from accretion disks, a higher temperature of the accretion disk at higher luminosity would harden the continuum.

### 3.2. Narrow Lyα Component

A unique feature of 3C 390.3 is its narrow–line variability. The optical narrow lines (FWHM $\leq 500$ km s$^{-1}$) have varied considerably over a period of more than 10 years (Zheng et al. 1995). This contrasts most other active galaxies in which the narrow–line intensities are stable (Antonucci & Cohen 1983) and widely serve as a robust flux calibrator. The most significant variation ($\sim 40\%$) of optical narrow lines was between 1983 September and 1984 April.

The flux of the fitted Lyα narrow component declined from $\sim 5 \times 10^{-14}$ in 1978 to $\sim 2 \times 10^{-14}$ erg cm$^{-2}$ s$^{-1}$ in 1984, and then remained approximately at a level of $3 \times 10^{-14}$ erg cm$^{-2}$ s$^{-1}$ since. In 1984, it varied in the period between 1984 June (SWP23179) and August (SWP23791), which is slightly earlier than the rapid variation in optical narrow lines (Zheng et al. 1995). During this period the continuum was at a minimum level. Another significant variation was observed between 1992 March and 1993 March, during which period the flux of narrow Lyα component decreased by $\sim 50\%$ (Figure 4).

### 3.3. Broad Components of Lyα Emission

While the peak positions of the blue and red components are assumed to match the humps in the Hα and Hβ emission profile, the humps in the Lyα and C IV profiles are visible only in several *IUE* spectra when 3C 390.3 was in high states (Figure 4). The measurements of SWP42367 find a velocity, with respect to the narrow peak, of $-4000 \pm 112$ km s$^{-1}$ for the blue component and $+5800 \pm 192$ km s$^{-1}$ for the red component. While these velocity displacements are in agreement with those of the optical counterparts, the blue component seems to be closer to the narrow peak.

The intensity of the red Ly$\alpha$ component is considerably weaker than the blue component. Note that the red Ly$\alpha$ component may contain a contribution from the N V emission which is indistinguishable in *IUE* spectra, therefore the intrinsic red Ly$\alpha$ component should be even weaker. Figure 2 plots the variations of the Ly$\alpha$ components.

### 3.4. C IV Emission

The total flux of C IV emission increases from $2.7 \times 10^{-13}$ erg cm$^{-2}$ s$^{-1}$ in SWP23791 to $1.3 \times 10^{-12}$ erg cm$^{-2}$ s$^{-1}$ in SWP42367. The variation scale of $4.9 \pm 0.5$ for C IV is very similar as that for Ly$\alpha$ ($4.8 \pm 0.5$). The comparable variation scale for Ly$\alpha$ and C IV in 3C 390.3 is different from that in many other Seyfert galaxies, such as Fairall 9 (Clavel, Wamsteker, & Glass 1989), in which the C IV flux varies significantly less than does Ly$\alpha$. The variations of the C IV components are plotted in Figure 3.

The C IV/Ly$\alpha$ ratio for the narrow component has remained at a level of $\sim 0.5$. For broad components, this ratio varies considerably. The averaged C IV/Ly$\alpha$ ratio is calculated for 3 subgroups of the *IUE* spectra that exhibit different flux levels: the high, medium and low states. Based on the total Ly$\alpha$ flux, each of the subgroups consists of 15, 14 and 14 *IUE* spectra, respectively. As shown in Table 4, the C IV/Ly$\alpha$ ratio for the central component is $\sim 0.2$. For the blue and red components, this ratio is $\sim 1.0$. Prominent differences in the C IV/Ly$\alpha$ ratio exist in individual spectra. In the spectrum SWP47089 (Figure 4), the blue Ly$\alpha$ component is stronger than the red while the red C IV component is stronger than the blue.

### 3.5. Cross–Correlation between Continuum and Emission Lines

The interpolated cross–correlation function (CCF, Gaskell & Peterson 1987; White & Peterson 1994) derives a time lag from two time series of events. Calculations are carried out between the continuum flux at 1335 Å and each of the fitted emission–line components. The results are listed in Table 5. The time lag $\Delta t_{peak}$ of the CCF yields the average distance between the BLR and the central power source as indicated by the maximum correlation coefficient $r_{max}$.

The narrow Ly$\alpha$ component does not show a correlation with the continuum. The poor temporal sampling rate and the low spectral resolution prohibit a reasonable estimate of the lag between the continuum and narrow Ly$\alpha$ component. The broad Ly$\alpha$ component is closely correlated with the continuum ($r \geq 0.74$), and the time lag of $\sim 50$ days is broadly consistent with that derived by Clavel & Wamsteker (1987).

It should be noted that the cross–correlation results shown in Table 5 depend heavily on the spectra obtained between 1991 and 1992 in which the continuum and lines were at maximum levels.

## 4. DISCUSSION

The peculiar double humps observed in the Hα and Hβ emission profiles in 3C 390.3 and their variations raise an intriguing question regarding the nature of the BLR in this object. The humps may be produced in a region dominated by Keplerian motion, preferably in an accretion disk (Pérez et al. 1988; Chen, Halpern, & Filippenko 1989; Eracleous & Halpern 1993). In general, the blue bump would be slightly stronger because of the Doppler boosting effect. More complicated models adopt a hot patch on the disk (Zheng et al. 1991), two-armed spiral shocks in accretion disks around central black holes (Chakrabarti & Wiita 1994) or nonspherical asymmetry (Eracleous et al. 1995) to account for the variable ratio of the red/blue hump.

On the other hand, double profile humps may be produced in a pair of conical line–emitting zones (Foltz, Wilkes, & Peterson 1983; Zheng et al. 1991, O'Brien, Goad, & Gondhalekar 1994). The radial velocity in each of the cones produces a distinct displacement between the two peaks, and the line width is explained in terms of angular extension of the cones. Such a double–stream model has been suggested to account for the observed profiles in IC 4329A (Marziani et al. 1992) and Ark 120 (Korista 1992). The Hα and Hβ emission features in these objects display a typical logarithmic profile, with a double peak separated by a few Å. The variation study in Ark 120 does not reveal any possible changes in the line profile and intensity along the profile wings.

Another model involves binary black holes (Gaskell 1983). If each of the black holes has its own broad–line region, the broad–line humps may be associated with the respective black hole. Precessing of the binary would result in a secular shift in the peak wavelength. Gaskell (1996) suggests a precessing period of $> 210$ years for 3C 390.3 and infers a black hole mass of $2 - 4 \times 10^9 M_\odot$.

In principle, these models may be tested with velocity–dependent line variations, i.e., with the time lag of different parts of the emission–line profile. In the case of a disk, both the blue and red wings would respond to the varying continuum with a similar time lag. In the case of a double outflow stream, the response to the continuum variation would develop from the blue wing, through the core to the red wing. A pattern of reversed order would indicate an inflow. In general, if the blue and red wing exhibit different time lags, the bipolar model would be in favor. The binary black–hole model estimates that the separation between the black holes are more than one lt-yr (Gaskell 1996), so any correlation between the flux variations of the two profile humps should be on the time scale of $\sim 1$ year.

Because Lyα and C IV emission are more sensitive to the physical conditions than the Balmer lines, and because the variations of UV lines are more prominent, the UV spectra provide clues to the formation of double–peaked profiles. In 3C 390.3, the Lyα profile wings appear to respond with a smaller time lag than the core, but the sampling has not been sufficient to distinguish with certainty the time lags for the broad–line components.

In addition, disk models predict that the two humps are quite symmetrical and, if they vary,

display a nearly sinusoidal pattern of variation. The *IUE* data indicate that (1) the blue component of Ly$\alpha$ is considerably stronger than the red component. The Doppler factor based on a velocity of $\sim 5000$ km s$^{-1}$ is not sufficient to produce a factor of two in the blue/red intensity ratio; (2) the blue component is nearly always stronger, except for two epochs in 1980 and 1992; and (3) the red/blue ratio varies in somewhat different ways for Ly$\alpha$ and C IV emission. In several epochs the red component is stronger than the blue for C IV, and vice versa for Ly$\alpha$. Such characteristics do not support the disk model.

It is natural to link the double–stream configuration with radio jet observed in this object. Because of the observed superluminal velocity, the jet should subtend an angle of less than $\theta \leq 1/\gamma$ to the line of sight where $\gamma$ is the Lorentz factor. Norman & Miley (1984) have proposed that the low–ionization lines may be formed in a jet. They argue that, as a jet propagates through and interacts with its environment via internal shocks and its unstable boundary layer cocoon, BLR clouds can be made in these interaction regions and sprayed off the jet into the general BLR. If the BLR in radio galaxies such as 3C 390.3 is associated with a jet, in the vicinity of a radio jet the shock wave may play a significant role in line formation.

The fitted narrow component shows a larger delay with the continuum of about 135 days. Still, it may not represent the true UV counterpart of the optical narrow lines that show secular variations. The spectrum SWP23791, obtained in 1984 August, shows a low state of the narrow components while the optical spectra taken in September (Zheng et al. 1995) show that the optical narrow lines were still in a high state, only in 1985 the optical spectra reveal the flux decline. Therefore it is likely that a significant portion of the UV narrow component is formed in the BLR.

## 5. SUMMARY

From the *IUE* spectra between 1978 and 1994, the following characteristics are found in 3C 390.3:

(1) The UV continuum varies by a factor of $\sim 27$ at 1350 Å and by $\sim 12$ at 1850 Å. Significant variations are observed within 30 days.

(2) The narrow–line components of Ly$\alpha$ and C IV vary by a factor of $\sim 2.5$. The broad components in profile wings vary with a time lag of $\sim 45$ days.

(3) The red Ly$\alpha$ component is generally weaker than the blue component except for several epochs. In 1993 March the red C IV component was considerably stronger than the blue component while the blue Ly$\alpha$ component was stronger.

These findings suggest that the observed extraordinary broad–line width in 3C 390.3 may not be from rotational motion. The differential variations between Ly$\alpha$ and C IV may be better explained in terms of a bipolar configuration.

The variability of narrow and broad lines in 3C 390.3 makes it a unique candidate to study the stratification of the line–emitting region. Monitoring campaigns at higher temporal frequency, high signal–to–noise level and higher spectral resolution will enable mapping of the velocity field in 3C 390.3 and reveal the nature of the double–peaked line profiles.

I thank B. M. Peterson for important comments and for providing his cross–correlation function code.

Fig. 1.— Variations of continuum flux in 3C 390.3.

Fig. 2.— Flux variations of fitted Ly$\alpha$ components in 3C 390.3. The narrow component is of width $\sim$ 9 Å. The blue component peak is at $\sim$ $-3600$ km s$^{-1}$ with respect to the narrow peak, and $\sim$ $+4500$ km s$^{-1}$ for the red component. All three broad components are assumed to have the same width.

Fig. 3.— Flux variations of fitted C IV components in 3C 390.3. The centroid wavelength and line width of each component are tied to the respective Ly$\alpha$ counterpart.

Fig. 4.— *IUE* spectra of 3C 390.3 and best–fit components. Note the change in continuum slope and a stronger red C IV component in SWP47089.

TABLE 1. Measurements of Continuum Flux in 3C 390.3[a]

TABLE 2. Measurements of Lyα Flux in 3C 390.3[a]

TABLE 3. Measurements of C IV Flux in 3C 390.3[a]

TABLE 4. Average C IV/Lyα Ratio for Emission–Line Components

TABLE 5. Cross–Correlation between Continuum and Fitted Broad–Line Components

Table 1.

| Year | Date | Julian Date (2440000+) | *IUE* Image | $F_{1360}$ | $F_{1850}$ |
|------|------|------------------------|-------------|------------|------------|
| 1978 | 325 | 3835 | SWP  3410 | 4.55 ± 0.26 | 2.33 ± 0.14 |
| 1978 | 332 | 3841 | SWP  3478 | 4.55 ± 0.51 | 2.10 ± 0.20 |
| 1979 | 047 | 3953 | SWP  4276 | 2.26 ± 0.17 | 1.76 ± 0.12 |
| 1979 | 092 | 3966 | SWP  4837 | 4.14 ± 0.45 | 2.03 ± 0.23 |
| 1979 | 096 | 3970 | SWP  4873 | 4.08 ± 0.32 | 2.23 ± 0.18 |
| 1980 | 147 | 4386 | SWP  9131 | 2.51 ± 0.24 | 1.32 ± 0.12 |
| 1980 | 339 | 4578 | SWP 10753 | 2.13 ± 0.18 | 1.92 ± 0.15 |
| 1980 | 362 | 4601 | SWP 10907 | 3.90 ± 0.17 | 2.37 ± 0.09 |
| 1980 | 031 | 4635 | SWP 11191 | 3.90 ± 0.51 | 2.37 ± 0.36 |
| 1982 | 207 | 5177 | SWP 17491 | 2.44 ± 0.33 | 1.50 ± 0.21 |
| 1982 | 311 | 5281 | SWP 18501 | 1.38 ± 0.62 | 0.94 ± 0.21 |
| 1983 | 087 | 5422 | SWP 19569 | 0.92 ± 0.26 | 0.58 ± 0.17 |
| 1983 | 266 | 5602 | SWP 21143 | 2.99 ± 0.30 | 1.47 ± 0.15 |
| 1984 | 090 | 5790 | SWP 22624 | 0.51 ± 0.23 | 1.40 ± 0.41 |
| 1984 | 157 | 5858 | SWP 23179 | 1.60 ± 0.20 | 0.85 ± 0.11 |
| 1984 | 240 | 5940 | SWP 23791 | 1.58 ± 0.32 | 1.01 ± 0.20 |
| 1985 | 085 | 6151 | SWP 25523 | 5.80 ± 0.20 | 2.63 ± 0.09 |
| 1986 | 242 | 6673 | SWP 29091 | 2.55 ± 0.23 | 2.14 ± 0.18 |
| 1987 | 028 | 6827 | SWP 30185 | 3.67 ± 0.35 | 1.71 ± 0.17 |
| 1987 | 156 | 6953 | SWP 31109 | 4.44 ± 0.29 | 1.37 ± 0.09 |
| 1987 | 224 | 7020 | SWP 31554 | 3.72 ± 0.20 | 3.24 ± 0.17 |
| 1987 | 365 | 7161 | SWP 32644 | 3.55 ± 0.48 | 2.75 ± 0.38 |
| 1988 | 047 | 7208 | SWP 32931 | 2.34 ± 0.18 | 1.48 ± 0.12 |
| 1989 | 081 | 7608 | SWP 35841 | 4.87 ± 0.23 | 2.49 ± 0.11 |
| 1989 | 129 | 7656 | SWP 36226 | 3.43 ± 0.18 | 1.51 ± 0.08 |
| 1989 | 226 | 7753 | SWP 36843 | 6.42 ± 0.27 | 3.94 ± 0.17 |
| 1989 | 317 | 7844 | SWP 37582 | 6.43 ± 0.30 | 4.85 ± 0.23 |
| 1990 | 029 | 7921 | SWP 38096 | 6.04 ± 0.36 | 2.96 ± 0.18 |
| 1990 | 048 | 7940 | SWP 38206 | 4.12 ± 0.14 | 2.36 ± 0.08 |
| 1990 | 107 | 7999 | SWP 38621 | 3.97 ± 0.26 | 2.92 ± 0.20 |
| 1990 | 192 | 8084 | SWP 39215 | 3.60 ± 0.12 | 1.80 ± 0.06 |
| 1990 | 242 | 8134 | SWP 39554 | 3.28 ± 0.12 | 2.03 ± 0.08 |
| 1990 | 244 | 8136 | SWP 39565 | 2.83 ± 0.14 | 1.83 ± 0.09 |
| 1990 | 362 | 8254 | SWP 40479 | 6.59 ± 0.26 | 4.43 ± 0.17 |

Table 1—Continued

| Year | Date | Julian Date (2440000+) | *IUE* Image | $F_{1360}$ | $F_{1850}$ |
|---|---|---|---|---|---|
| 1991 | 113 | 8370 | SWP 41459 | 6.19 ± 0.17 | 3.65 ± 0.11 |
| 1991 | 243 | 8500 | SWP 42367 | 3.50 ± 0.29 | 7.16 ± 0.15 |
| 1992 | 013 | 8635 | SWP 43634 | 9.20 ± 0.27 | 4.27 ± 0.14 |
| 1992 | 040 | 8662 | SWP 43951 | 5.22 ± 0.14 | 3.22 ± 0.09 |
| 1992 | 141 | 8763 | SWP 44725 | 5.22 ± 0.24 | 3.22 ± 0.15 |
| 1993 | 061 | 9049 | SWP 47089 | 2.17 ± 0.21 | 1.34 ± 0.14 |
| 1993 | 154 | 9143 | SWP 47799 | 3.34 ± 0.38 | 2.81 ± 0.32 |
| 1993 | 259 | 9247 | SWP 48650 | 8.01 ± 0.42 | 4.38 ± 0.23 |
| 1994 | 168 | 9521 | SWP 51116 | 3.16 ± 0.39 | 2.20 ± 0.11 |

[a]Continuum flux in units of $10^{-15}\,\mathrm{erg\,cm^{-2}\,s^{-1}\,Å^{-1}}$, derived from the best–fit power continuum for given wavelengths.

– 16 –Table 2.

| Julian Date (2440000+) | $F_{narrow}$ | $F_{blue}$ | $F_{red}$ | $F_{central}$ |
|---|---|---|---|---|
| 3835 | 4.51 ± 0.19 | 0.91 ± 0.20 | 0.79 ± 0.30 | 7.13 ± 0.08 |
| 3841 | 5.23 ± 0.11 | 2.95 ± 0.67 | 3.77 ± 1.69 | 5.48 ± 0.74 |
| 3953 | 4.40 ± 0.32 | 1.95 ± 0.37 | 0.92 ± 0.46 | 6.71 ± 0.97 |
| 3966 | 5.77 ± 0.83 | 1.29 ± 0.78 | 1.71 ± 0.66 | 3.65 ± 1.30 |
| 3970 | 4.72 ± 0.84 | 2.11 ± 0.71 | 2.06 ± 0.38 | 4.42 ± 0.77 |
| 4386 | 3.84 ± 0.24 | 1.17 ± 0.25 | 0.67 ± 0.24 | 5.75 ± 0.39 |
| 4578 | 3.69 ± 0.07 | 1.31 ± 0.24 | 1.33 ± 0.26 | 4.88 ± 0.19 |
| 4601 | 3.82 ± 0.07 | 1.37 ± 0.19 | 1.73 ± 0.23 | 4.46 ± 0.43 |
| 4635 | 3.57 ± 0.34 | 1.82 ± 0.34 | 0.00 | 6.60 ± 0.68 |
| 5177 | 3.63 ± 0.46 | 1.63 ± 0.50 | 1.29 ± 0.31 | 1.99 ± 1.06 |
| 5281 | 3.50 ± 0.32 | 1.89 ± 0.25 | 0.56 ± 0.26 | 2.74 ± 0.45 |
| 5422 | 3.44 ± 0.34 | 1.32 ± 0.30 | 0.56 ± 0.28 | 3.50 ± 0.86 |
| 5602 | 3.19 ± 0.29 | 1.95 ± 0.28 | 1.24 ± 0.59 | 5.38 ± 1.62 |
| 5790 | 3.73 ± 0.64 | 1.71 ± 0.65 | 1.31 ± 0.65 | 2.32 ± 0.85 |
| 5858 | 3.76 ± 0.29 | 1.42 ± 0.18 | 1.09 ± 0.13 | 2.06 ± 0.47 |
| 5940 | 1.92 ± 0.65 | 0.00 | 0.09 ± 0.35 | 2.30 ± 0.38 |
| 6151 | 3.72 ± 0.07 | 3.70 ± 0.20 | 3.11 ± 0.26 | 2.82 ± 0.27 |
| 6673 | 2.92 ± 0.24 | 1.47 ± 0.35 | 0.15 ± 0.72 | 6.21 ± 0.41 |
| 6824 | 2.88 ± 0.56 | 2.49 ± 0.56 | 1.59 ± 0.40 | 4.95 ± 1.41 |
| 6953 | 2.43 ± 0.58 | 2.43 ± 0.20 | 1.13 ± 0.14 | 4.91 ± 0.82 |
| 7020 | 2.68 ± 0.46 | 1.46 ± 0.22 | 0.36 ± 0.20 | 6.64 ± 0.57 |
| 7161 | 2.48 ± 0.28 | 2.13 ± 0.86 | 0.45 ± 0.21 | 5.01 ± 0.83 |
| 7208 | 2.31 ± 0.18 | 2.43 ± 0.25 | 1.28 ± 0.17 | 4.30 ± 0.41 |
| 7608 | 2.39 ± 0.42 | 3.91 ± 0.51 | 2.30 ± 0.26 | 7.08 ± 1.07 |
| 7656 | 2.44 ± 0.22 | 3.28 ± 0.36 | 1.49 ± 0.46 | 7.24 ± 0.52 |
| 7753 | 2.69 ± 0.31 | 4.10 ± 0.41 | 2.03 ± 0.16 | 6.36 ± 0.77 |
| 7844 | 3.02 ± 0.34 | 4.34 ± 0.39 | 1.89 ± 0.26 | 9.30 ± 0.15 |
| 7921 | 2.21 ± 0.88 | 4.23 ± 1.52 | 1.59 ± 0.75 | 7.95 ± 2.83 |
| 7940 | 2.46 ± 0.22 | 2.96 ± 1.90 | 2.34 ± 0.26 | 8.93 ± 1.96 |
| 7999 | 2.23 ± 0.40 | 3.77 ± 0.36 | 1.59 ± 0.25 | 6.61 ± 0.77 |
| 8084 | 3.23 ± 0.16 | 3.60 ± 1.06 | 1.58 ± 0.18 | 4.99 ± 1.05 |
| 8134 | 2.68 ± 0.40 | 2.94 ± 0.45 | 1.16 ± 0.12 | 5.39 ± 0.80 |
| 8136 | 2.51 ± 0.29 | 2.52 ± 0.28 | 0.94 ± 0.23 | 5.77 ± 0.56 |
| 8254 | 2.86 ± 0.40 | 3.05 ± 0.31 | 1.30 ± 0.34 | 5.95 ± 0.32 |

Table 2—Continued

| Julian Date (2440000+) | $F_{narrow}$ | $F_{blue}$ | $F_{red}$ | $F_{central}$ |
|---|---|---|---|---|
| 8370 | 2.35 ± 0.31 | 3.23 ± 0.23 | 1.87 ± 0.17 | 7.59 ± 0.61 |
| 8500 | 2.20 ± 0.18 | 4.23 ± 1.09 | 3.15 ± 0.41 | 9.95 ± 1.45 |
| 8635 | 2.65 ± 1.03 | 5.45 ± 0.70 | 3.69 ± 0.25 | 10.77± 2.24 |
| 8662 | 1.97 ± 0.30 | 5.39 ± 0.24 | 3.57 ± 0.19 | 10.54± 0.55 |
| 8763 | 2.23 ± 0.37 | 5.13 ± 0.26 | 2.24 ± 0.21 | 9.23 ± 0.75 |
| 9049 | 1.22 ± 0.13 | 1.58 ± 0.68 | 0.00 | 5.32 ± 0.82 |
| 9143 | 2.51 ± 0.49 | 1.82 ± 0.29 | 1.37 ± 0.18 | 5.07 ± 0.93 |
| 9247 | 2.10 ± 0.70 | 2.50 ± 0.27 | 2.15 ± 0.21 | 6.40 ± 1.05 |
| 9521 | 2.60 ± 0.27 | 2.22 ± 0.94 | 0.48 ± 0.67 | 3.77 ± 2.22 |

[a]Line flux in units of $10^{-13}$ erg cm$^{-2}$ s$^{-1}$.

Table 3.

| Julian Date (2440000+) | $F_{narrow}$ | $F_{blue}$ | $F_{red}$ | $F_{central}$ |
|---|---|---|---|---|
| 3835 | 2.51 ± 0.15 | 2.39 ± 0.43 | 0.62 ± 0.24 | 1.86 ± 0.77 |
| 3841 | 2.07 ± 0.83 | 3.31 ± 1.08 | 2.33 ± 0.40 | 3.94 ± 0.52 |
| 3953 | 2.29 ± 0.17 | 2.02 ± 0.45 | 2.76 ± 0.46 | 1.18 ± 0.94 |
| 3966 | 1.70 ± 0.22 | 3.19 ± 0.81 | 2.27 ± 0.40 | 1.78 ± 0.41 |
| 3970 | 2.72 ± 0.19 | 4.38 ± 1.26 | 2.00 ± 0.36 | 0.04 ± 0.92 |
| 4386 | 2.21 ± 0.11 | 1.75 ± 0.23 | 1.29 ± 0.21 | 0.00 |
| 4578 | 1.90 ± 0.12 | 1.56 ± 0.14 | 1.76 ± 0.29 | 0.45 ± 0.17 |
| 4601 | 1.57 ± 0.11 | 1.07 ± 0.17 | 2.04 ± 0.37 | 0.82 ± 0.62 |
| 4635 | 2.01 ± 0.26 | 2.76 ± 0.74 | 1.90 ± 0.42 | 0.86 ± 1.26 |
| 5177 | 1.50 ± 0.37 | 0.39 ± 0.33 | 0.10 ± 0.24 | 0.37 ± 0.69 |
| 5281 | 1.42 ± 0.13 | 0.61 ± 0.27 | 0.83 ± 0.24 | 0.00 |
| 5422 | 1.34 ± 0.16 | 0.69 ± 0.24 | 1.21 ± 0.17 | 0.24 ± 0.44 |
| 5602 | 1.22 ± 0.04 | 1.06 ± 0.61 | 1.28 ± 0.25 | 1.85 ± 1.04 |
| 5790 | 1.60 ± 0.28 | 1.49 ± 0.17 | 0.88 ± 0.39 | 0.19 ± 0.78 |
| 5858 | 1.34 ± 0.08 | 0.85 ± 0.13 | 0.65 ± 0.13 | 0.00 |
| 5940 | 1.02 ± 0.14 | 1.70 ± 0.52 | 0.00 | 0.00 |
| 6151 | 1.34 ± 0.44 | 2.81 ± 0.84 | 2.75 ± 0.42 | 1.20 ± 0.37 |
| 6673 | 0.65 ± 0.14 | 1.71 ± 0.24 | 1.46 ± 0.12 | 0.70 ± 0.33 |
| 6824 | 1.30 ± 0.25 | 3.01 ± 1.08 | 2.49 ± 0.42 | 0.33 ± 1.56 |
| 6953 | 1.12 ± 0.17 | 0.83 ± 0.22 | 1.38 ± 0.14 | 1.42 ± 0.34 |
| 7020 | 1.22 ± 0.17 | 2.20 ± 0.47 | 1.76 ± 0.21 | 1.47 ± 0.66 |
| 7161 | 1.57 ± 0.17 | 1.58 ± 0.73 | 1.32 ± 0.19 | 0.82 ± 0.93 |
| 7208 | 1.28 ± 0.12 | 2.81 ± 0.18 | 2.12 ± 0.14 | 0.00 |
| 7608 | 1.26 ± 0.23 | 4.85 ± 0.80 | 2.81 ± 0.21 | 1.91 ± 1.01 |
| 7656 | 1.11 ± 0.11 | 3.90 ± 0.46 | 2.02 ± 0.23 | 2.40 ± 0.29 |
| 7753 | 0.92 ± 0.55 | 3.65 ± 0.58 | 2.12 ± 0.12 | 3.50 ± 1.00 |
| 7844 | 1.08 ± 0.16 | 3.72 ± 0.17 | 2.19 ± 0.23 | 4.48 ± 0.56 |
| 7921 | 0.89 ± 0.30 | 3.53 ± 1.06 | 1.29 ± 0.98 | 3.57 ± 1.34 |
| 7940 | 1.20 ± 0.06 | 5.07 ± 1.57 | 2.82 ± 0.20 | 3.68 ± 1.88 |
| 7999 | 0.88 ± 0.21 | 2.14 ± 0.39 | 1.70 ± 0.24 | 3.11 ± 0.58 |
| 8084 | 1.23 ± 0.12 | 2.76 ± 0.67 | 1.56 ± 0.17 | 2.61 ± 0.72 |
| 8134 | 0.80 ± 0.09 | 1.57 ± 0.36 | 1.12 ± 0.13 | 1.75 ± 0.38 |
| 8136 | 0.97 ± 0.13 | 2.17 ± 0.57 | 1.13 ± 0.17 | 1.22 ± 0.63 |
| 8254 | 1.62 ± 0.11 | 2.58 ± 0.90 | 2.87 ± 0.21 | 1.15 ± 1.19 |

Table 3—Continued

| Julian Date (2440000+) | $F_{\mathrm{narrow}}$ | $F_{\mathrm{blue}}$ | $F_{\mathrm{red}}$ | $F_{\mathrm{central}}$ |
|---|---|---|---|---|
| 8370 | 1.02 ± 0.16 | 3.54 ± 0.22 | 2.29 ± 0.17 | 2.96 ± 0.45 |
| 8500 | 1.41 ± 0.15 | 4.91 ± 0.21 | 2.92 ± 0.19 | 5.05 ± 0.40 |
| 8635 | 1.56 ± 0.50 | 6.95 ± 0.61 | 4.98 ± 0.17 | 4.04 ± 1.60 |
| 8662 | 1.22 ± 0.16 | 5.78 ± 0.23 | 3.45 ± 0.18 | 5.40 ± 0.36 |
| 8763 | 1.15 ± 0.16 | 4.63 ± 0.36 | 2.94 ± 0.17 | 4.03 ± 0.56 |
| 9049 | 0.70 ± 0.08 | 1.30 ± 0.29 | 3.43 ± 0.10 | 0.24 ± 0.12 |
| 9143 | 1.41 ± 0.18 | 0.88 ± 0.33 | 0.34 ± 0.16 | 1.41 ± 0.49 |
| 9247 | 1.51 ± 0.29 | 2.57 ± 0.34 | 2.54 ± 0.20 | 1.64 ± 0.64 |
| 9521 | 0.87 ± 0.07 | 1.32 ± 0.62 | 0.14 ± 0.20 | 0.81 ± 0.88 |

[a]Line flux in units of $10^{-13}\,\mathrm{erg\,cm^{-2}\,s^{-1}}$.

Table 4.

|  | Narrow | Blue | Red | Central |
|---|---|---|---|---|
| High States | 0.46 ± 0.21 | 0.87 ± 0.57 | 1.14 ± 0.88 | 0.34 ± 0.19 |
| Medium States | 0.45 ± 0.24 | 0.99 ± 0.68 | 1.27 ± 0.64 | 0.14 ± 0.10 |
| Low States | 0.40 ± 0.11 | 0.65 ± 0.46 | 0.72 ± 0.64 | 0.15 ± 0.16 |

Table 5.

| Component | $\Delta t_{peak}$ (day) | FWHM (day) | $r_{max}$ |
|---|---|---|---|
| Ly$\alpha$ central | 66 | 235 | 0.81 |
| Ly$\alpha$ blue | 46 | 231 | 0.77 |
| Ly$\alpha$ red | 42 | 200 | 0.74 |
| Ly$\alpha$ total | 66 | 221 | 0.83 |
| C IV central | 53 | 230 | 0.78 |
| C IV blue | 42 | 216 | 0.76 |
| C IV red | 58 | 170 | 0.71 |
| C IV total | 65 | 214 | 0.84 |

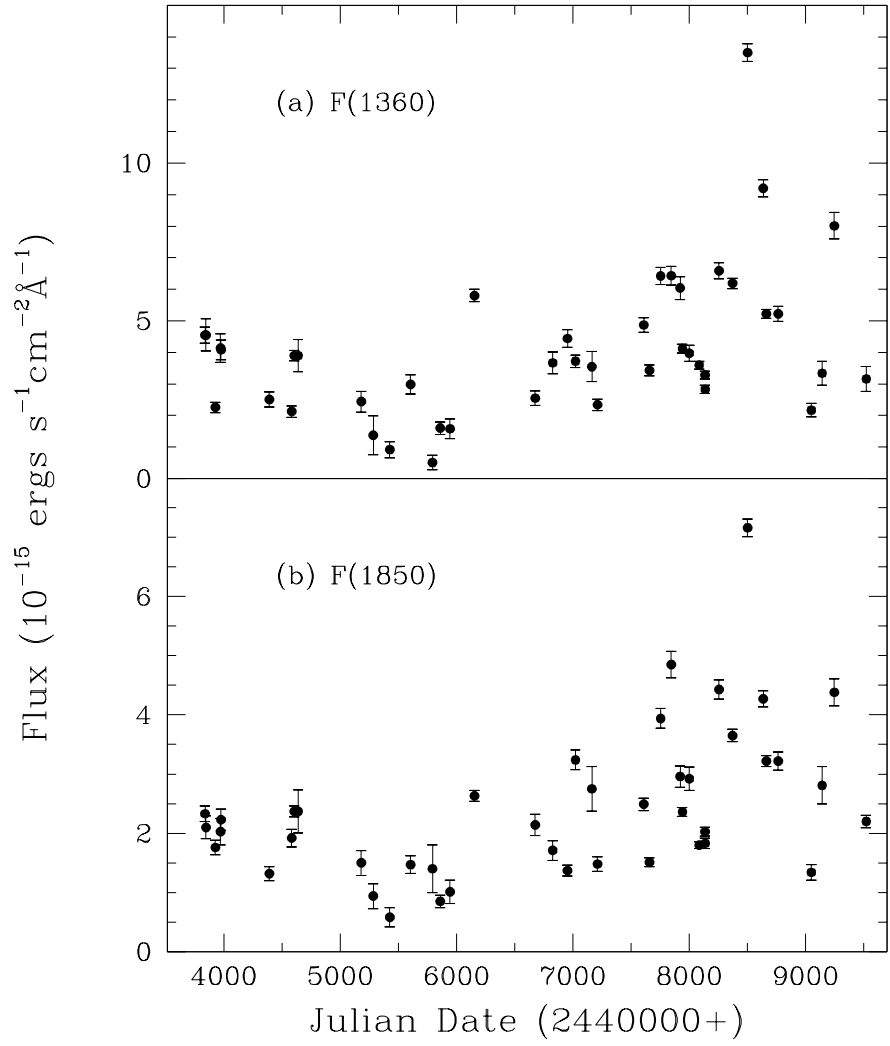

Fig. 1.—

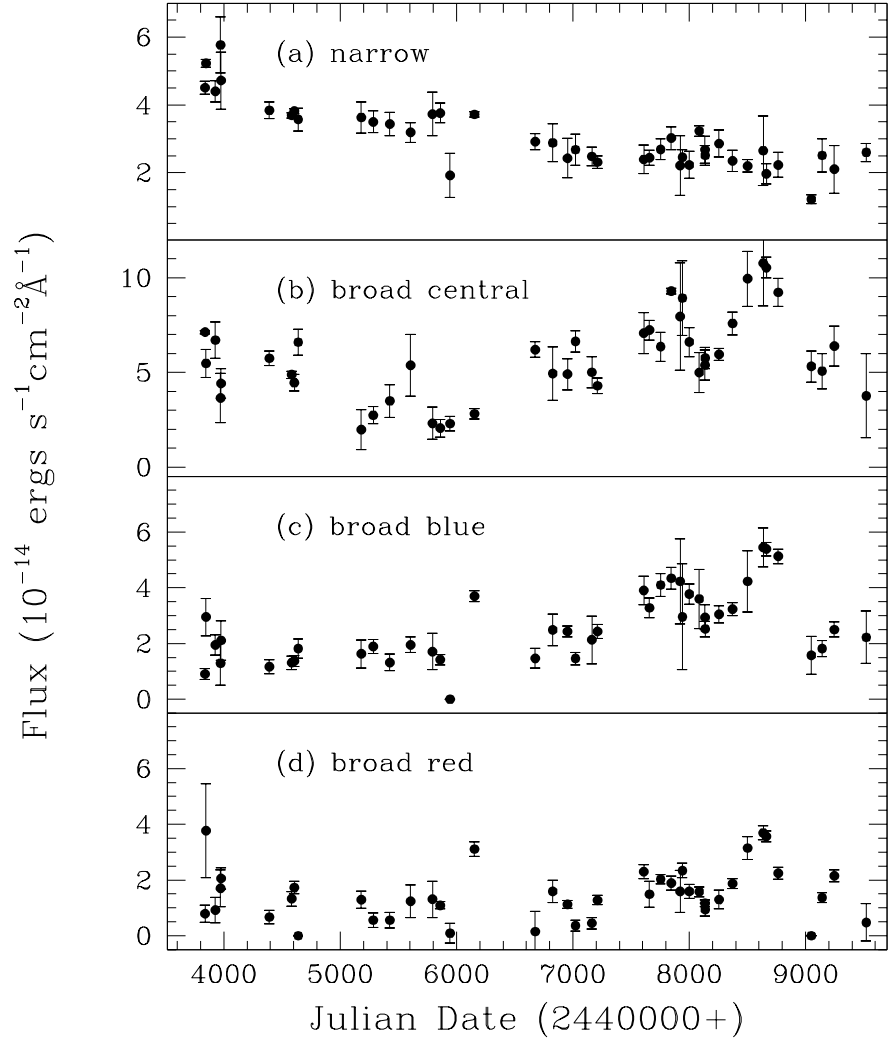

Fig. 2.—

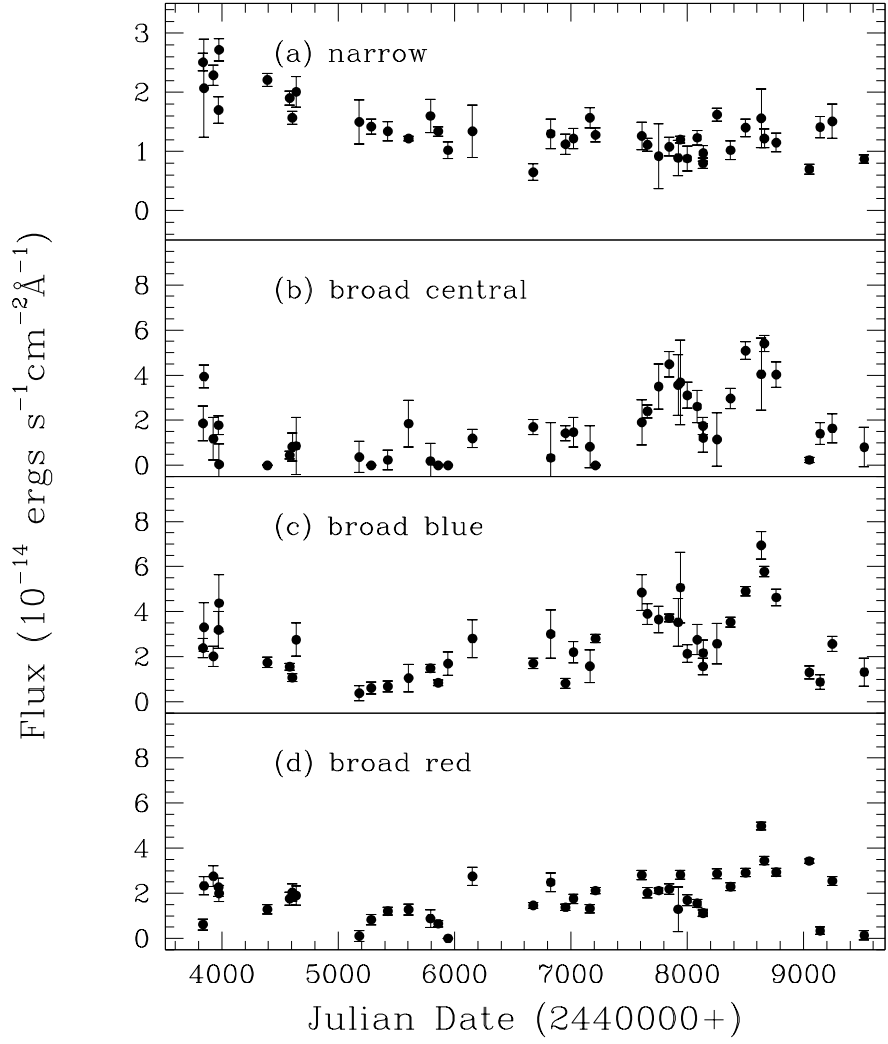

Fig. 3.—

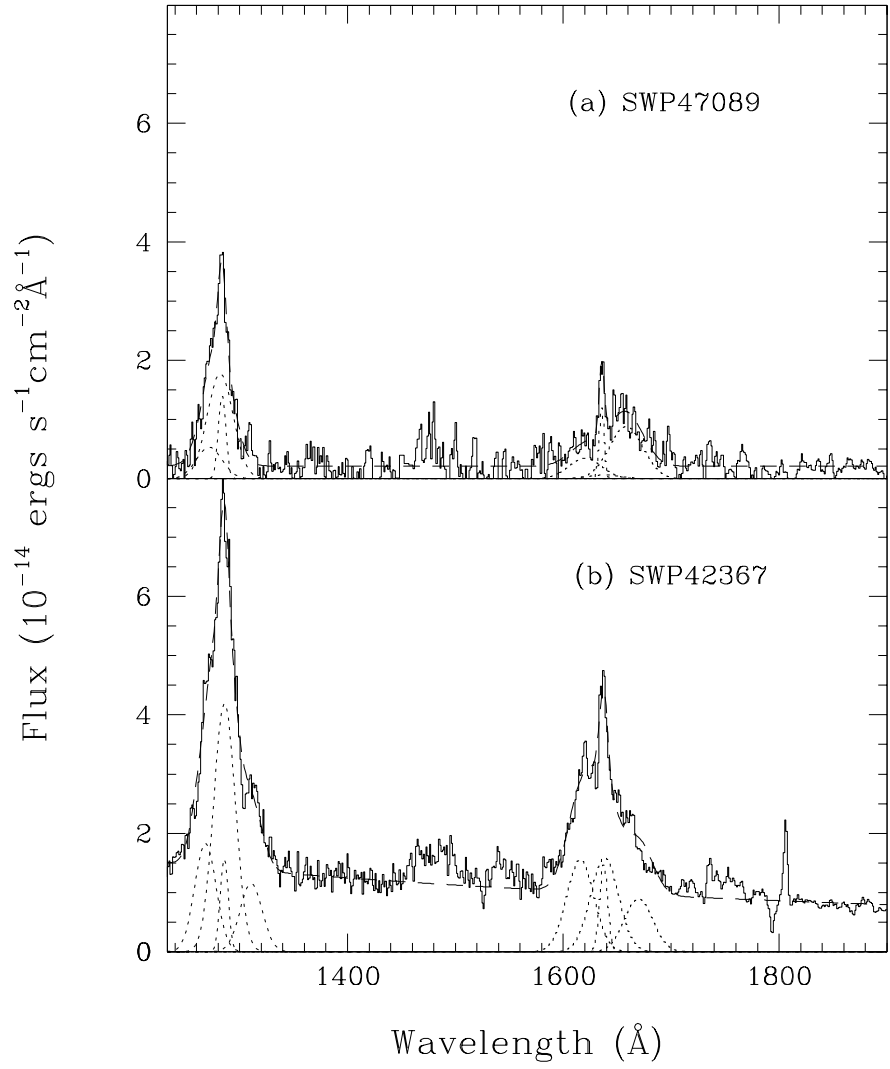

Fig. 4.—